# Comment: Microarrays, Empirical Bayes and the Two-Groups Model

**Kenneth Rice and David Spiegelhalter**

Through his various examples, Professor Efron makes a convincing case that cutting-edge science requires methods for detecting multiple "non-nulls." These methods must be straightforward to implement, but perhaps more importantly statisticians need to be able to justify them unambiguously. Efron's Empirical Bayes approach is certainly computationally efficient, but we feel the rationale for making each of his steps is unattractively ad hoc. This concern is practical, not philosophical; Efron's criterion for choice of tuning parameters seems to be that they look "believable." In less expert hands, this approach seems to introduce a lot of leeway for practitioners to simply "tune" away until they get the results they want.

In an attempt to address this problem, we will describe an approach developed in a fully model-based framework. As with locfdr, the calculations are fast, but our whole analysis derives from clear up-front statements about what the analysis is trying to achieve, and the modeling assumptions made. The results look reassuringly similar to Professor Efron's. We hope this will be helpful for understanding the current paper, and in making a contribution to this general field.

We begin by following Efron in placing the local false discovery rate, $\mathrm{fdr}(z)$, as the primary focus of the analysis, and exploit the fact that it can offer a neat parameterization of the two-part model. If the

marginal, "mixture" density for the $z$-values is

$$f(z) = p_0 f_0(z) + (1 - p_0) f_1(z)$$

and $\mathrm{fdr}(z) = p_0 f_0(z)/f(z)$, then

$$f_1(z) = \frac{p_0}{1 - p_0} \frac{1 - \mathrm{fdr}(z)}{\mathrm{fdr}(z)} f_0(z).$$

We observe that, because $f_1$ is a density, we only need to know $f_0$ and fdr in order to find its normalized form, and in turn this tells us the value of $p_0$. Thus, for a given $f_0$, specifying fdr sets up everything else we require for model-based analysis.

Naturally, the analysis we report will depend on the functional form assumed for fdr, and Efron implicitly assumes a rather flexible form of fdr, through a seventh-order polynomial-smoothed density estimate. However, this approach does not rule out an $\widehat{\mathrm{fdr}}$ with multiple peaks. Thinking of the schools example, we would not want to be the statistician explaining how two "bad" schools may have $z_1 < z_2 < 0$, but yet $\widehat{\mathrm{fdr}}(z_1) > 0.2$ while $\widehat{\mathrm{fdr}}(z_2) < 0.2$. Put more simply, Efron's method can report that School 1 has worse performance, but only School 2 is called an outlier. We find it more straightforward to a priori justify our choice of fdr by careful consideration of its role in the reported inference.

In our experience, the search for non-null "discoveries" is based around two ideas; first, we will not discover anything near the center of $f_0$ (effectively Efron's "zero assumption," also termed "purity" by Genovese and Wasserman, 2004). A second sensible assumption is that the evidence for $z$ being "null" will decrease monotonically as we move out from the center. One way to satisfy this is with a logistic-linear form for fdr, giving a two-component normal mixture for $f_1$, but we get closer to the spirit of Efron's analysis by assuming that fdr is unity inside a central region, and then follows a half-normal decline, that is,

$$\mathrm{fdr}^H(z) = \begin{cases} e^{-(z+k_a)^2/2}, & z < -k_a, \\ 1, & -k_a \le z \le k_b, \\ e^{-(z-k_b)^2/2}, & z > k_b. \end{cases}$$

*Kenneth Rice is Assistant Professor, Department of Biostatistics, University of Washington, Seattle, Washington 98195-7232, USA (e-mail: kenrice@u.washington.edu). David Spiegelhalter is Senior Scientist, MRC Biostatistics Unit, Institute of Public Health, Cambridge CB2 OSR, United Kingdom (e-mail: David.Spiegelhalter@mrc-bsu.cam.ac.uk).*

Following the observation above, taking the null component $f_0$ to be standard Normal, now defines the following marginal distribution $f^H(z)$:

$$f^H(z) = p_0 (2\pi)^{-1/2} \cdot \begin{cases} e^{-|z| k_a + k_a^2/2}, & z < -k_a, \\ e^{-z^2/2}, & -k_a \leq z \leq k_b, \\ e^{-|z| k_b + k_b^2/2}, & z > k_b, \end{cases}$$

where the constant of proportionality is $p_0$, the proportion of nulls, which is an easily determined function of $k_a$ and $k_b$.

$f^H(z)$ is seen to have a $N(0,1)$ "core" and exponential tails. By substituting $(z - \mu_0)/\sigma_0$ for $z$ in $f_0(z)$ and $\text{fdr}^H(z)$, it is easily generalized to a full location-scale family, where the "core" (or null distribution) is now $N(\mu_0, \sigma_0^2)$. We term this a "Huber" distribution, denoted $H(\mu_0, \sigma_0, k_a, k_b)$, following the observation in Huber (1964) that his optimal robust location estimation procedure based on a piecewise-linear bounded influence function was precisely equivalent to maximum likelihood estimation applied to such a distribution, but with $k_a = k_b = k$ specified and $\sigma_0$ assumed known.

Assuming this distribution and adopting a full likelihood approach, maximum likelihood estimates $\hat{\mu}_0, \hat{\sigma}_0$ are the solutions of estimating equations that take, up to a very good approximation, the same form as Huber's famous "Type 2" estimator. We do not need to fix $k_a$ and $k_b$; they can be estimated from the data in the same way.

We have implemented maximum likelihood-based regression for this error distribution within our own R package (*huber.lm*), and also as a fully Bayesian MCMC approach via a new distribution, *dhuber*, within WinBUGS.

Figure 1 and Table 1 show the results of fitting this distributional family to four of Efron's examples using *huber.lm*.

In line with Efron, we assume that $f_0$ follows a $N(\mu_0, \sigma_0^2)$ distribution, and provide point estimates for $\mu_0, \sigma_0, p_0$ as well as $k_a, k_b$. We also show the fitted marginal distributions $f^H(z)$, QQ-plots of the $z$-values against $f^H(z)$ and a "naive" Normal, the fitted local false discovery rate $\text{fdr}^H(z)$, and an appropriately scaled representation of the "alternative" distribution $f_1$. Figure 1 shows a good fit of the Huber distribution to these examples. The fitted $\text{fdr}^H$ curves are also plotted, and these show a close concordance with Efron's locfdr results. For the BRCA data, we have not plotted $\text{fdr}^H$, as use of the Huber distribution here gives estimates for both $k_a$ and $k_b$ tending to $\infty$, and hence gives a point estimate of

$\widehat{\text{fdr}} = 1$ for all data points. The practical message is clear; we find that the BRCA data, on its own, provides no strong evidence of any signals beyond the fitted $N(\mu, \sigma^2)$ null, in line with Efron's results. The QQ-plot for the BRCA data provides further informal confirmation. Other authors have declared some evidence for signals in this dataset, a recent example being Jin and Cai (2007). However, this is in contrast to a Bayesian analysis with a uniform prior for $k_a$ and $k_b$, which leads to a posterior for both $k_a$ and $k_b$ that rules out values less than 2 ($p_0 > 0.8\%$) and which provides an essentially uniform distribution for $k_a, k_b > 3$ ($p_0 < 0.02\%$).

Table 1 provides parameter estimates for the asymmetric Huber distribution: likelihood ratio tests for common $k$ are $p = 0.68$ (Prostate); $p = 0.14$ (Education); $p = 0.007$ (HIV). We find a close concordance between our results and those in Efron's paper. The estimated proportions of nonnull observations are 1.7% (Prostate), 7.3% (Education) and 6.2% (HIV). As $p_0$ is a slightly messy function of $k_a$ and $k_b$,

$$p_0 = \sqrt{2\pi}[e^{-k_a^2/2}/k_a + e^{-k_b^2/2}/k_b + \sqrt{2\pi}(\Phi(k_a) + \Phi(k_b) - 1)]^{-1},$$

we have found it easiest to obtain intervals by using an MCMC approach. However, using the delta method or a parametric bootstrap on the distribution of the MLEs offers, in spirit, the same inference.

In contrast to Efron's desire to "minimize the amount of statistical modeling required of the statistician," we would encourage statistical modeling *where the modeling assumptions are clear and comprehensible*; for example, we find a simply defined parametric model preferable to Efron's seven-parameter polynomial-smoothed density estimate. Our explicit acknowledgment of these assumptions also motivates consideration (below) of how they may be usefully strengthened, and also whether they may be relaxed.

Using a simple but flexible fully parametric family such as the Huber distributions confers many advantages. If we are willing to condition on the adequacy of the assumed model for $f^H(z)$, then the full resources of likelihood modeling become available, providing interval estimates, hypothesis tests and so on. In a hierarchical setting, the Huber distribution can also be considered at the random-effects level. Computationally this is handled with ease within a full Bayesian MCMC environment, where using $H(\mu, \sigma, k)$ or $H(\mu, \sigma, k_a, k_b)$ within a hierarchical model presents no additional difficulties over its use as a sampling distribution. Becoming "more" Bayesian still, we note the possibilities for use of informative priors regarding the thresholds $k_a$ and $k_b$, and hence implicitly $p_0$. In our opinion, analyses which acknowledge these a priori assumptions seem particularly attractive for examples smaller than Efron's, where a reliable density estimate seems out of reach. Finally, a Bayesian modeling framework allows the inclusion of a model for such data within an integrated evidence synthesis, which can be guided by a combination of substantive knowledge and data analysis.

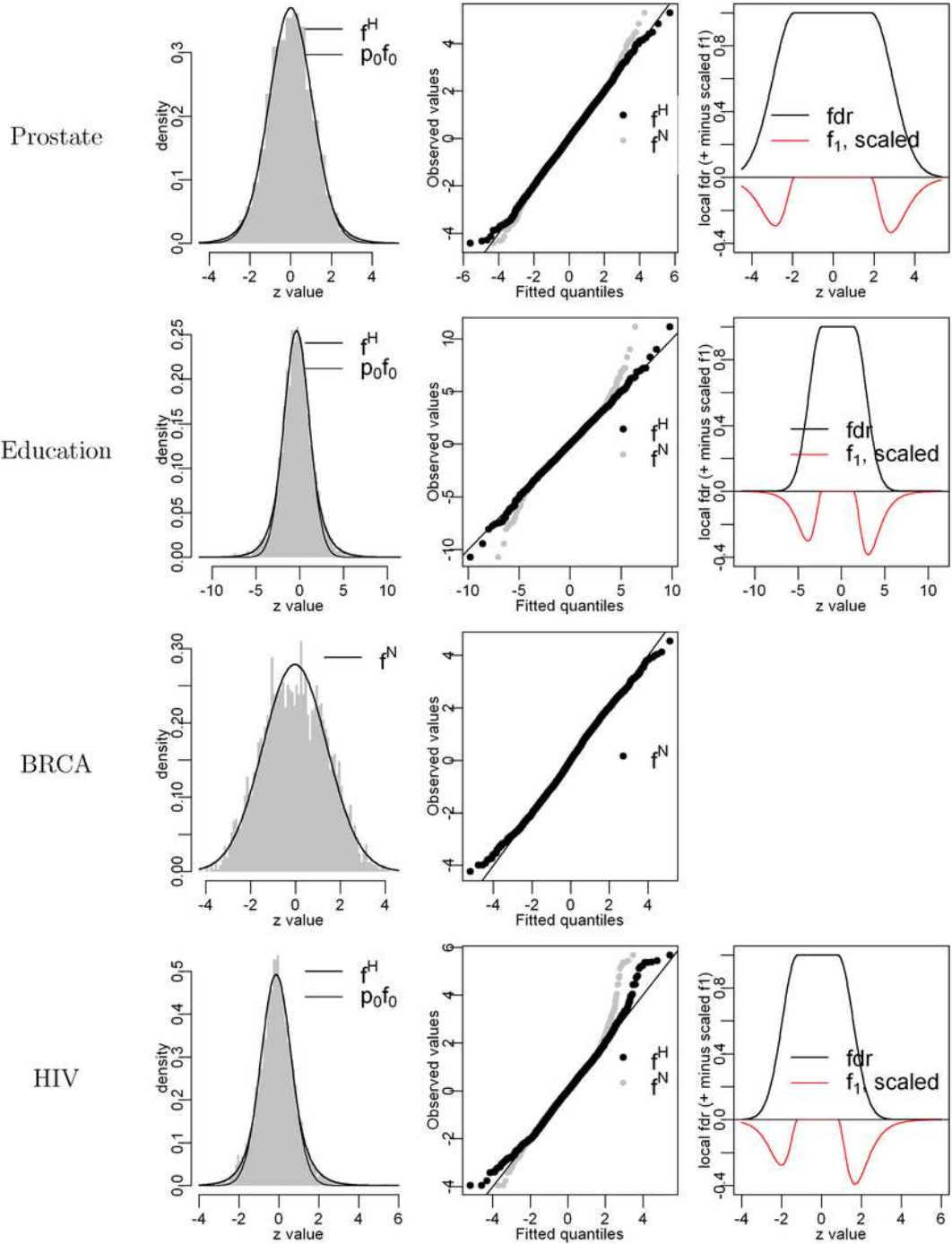

FIG. 1. *Summary plots fitting the Huber distribution to four examples. For each dataset, we plot histograms of the z-values and fitted marginal distribution, QQ-plots of the data against fitted Huber distribution ($f^H$) and a naive pure Normal ($f^N$), and finally a plot of the fitted fdr and the alternative distribution $f_1$ (inverted). For BRCA, the fitted fdr is always 1, giving no strong evidence of signals in this dataset.*

TABLE 1
*Maximum likelihood estimates (95% intervals) for parameters of the asymmetric Huber distribution for four of Efron's examples; the intervals for $p_0$ are obtained from an MCMC simulation*

|  | Prostate | | Education | | BRCA | | HIV | |
|---|---|---|---|---|---|---|---|---|
| $\mu_0$ | −0.001 | (−0.031, 0.030) | −0.361 | (−0.427, −0.295) | −0.026 | (−0.075, 0.023) | −0.138 | (−0.161, −0.115) |
| $\sigma_0$ | 1.059 | (1.030, 1.089) | 1.452 | (1.363, 1.546) | 1.431 | (1.396, 1.466) | 0.760 | (0.730, 0.791) |
| $k_a$ | 1.80 | (1.61, 2.01) | 1.31 | (1.17, 1.48) | — | — | 1.40 | (1.28, 1.53) |
| $k_b$ | 1.75 | (1.59, 1.93) | 1.21 | (1.08, 1.37) | — | — | 1.26 | (1.17, 1.36) |
| $p_0$ | 0.983 | (0.975, 0.990) | 0.927 | (0.899, 0.950) | — | — | 0.938 | (0.921, 0.954) |

Taking a less Bayesian or full-likelihood approach, and not wishing to condition on the "truth" of the model assumptions, one could proceed directly to Huber-style estimating equations for $\mu_0, \sigma_0$ and $k$ (or $k_a$ and $k_b$), justified either through their connection to the model we have described, or by arguing that this influence function directly reflects the population parameter we want to estimate; if we are trying to minimize model-dependence, the second approach is more satisfactory, and is quite standard in GEE. Sandwich and/or bootstrap variance estimates could be used to reflect uncertainty about these point estimates, without further parametric assumptions about the mixture distribution $f$. In samples of thousands of $z$'s (but not with a few hundred), this provides appealingly robust estimates of location and scale.

However, going beyond $\mu_0$ and $\sigma_0$, it is not clear to us that the GEE paradigm allows "model-robust" measures of fdr. Must one compare the marginal $f$ to an $f_0$ which is assumed to have a specifically Gaussian form, or that of some other parametric family? Might some advanced form of cross-validation offer a model-free approach? And could this be done without an excessive computational burden? Any insights from Professor Efron in this matter would be very welcome.

In conclusion, we feel that flexible likelihood or Bayesian modeling techniques, combined with basic insights from the literature on outlier-robustness, will contain much of value in the era of microarrays and other data-sources requiring large numbers of hypothesis tests. We thank Professor Efron for his stimulating paper, and also for his generosity in making available the four featured datasets.